\documentclass[a4paper,11pt]{article}
\usepackage[utf8]{inputenc}
\usepackage{setspace}
\usepackage{epsfig}
\usepackage{amsfonts}
\usepackage{amssymb}
\usepackage{amsmath}
\usepackage{subfigure}
\usepackage[usenames]{color}
\usepackage{comment}
\usepackage{multirow}
\usepackage{psfrag}
\usepackage{fullpage}
\usepackage{verbatim} 
\usepackage{soul}

\begin{document}

\title{IEEE 802.11ax: High-Efficiency WLANs\footnote{Accepted for publication in IEEE Wireless Communications Magazine. July 2015}}
\author{Boris Bellalta\\ Universitat Pompeu Fabra, Barcelona}

\date{}

\maketitle

\onehalfspace

\begin{abstract} 
IEEE 802.11ax-2019 will replace both IEEE 802.11n-2009 and IEEE 802.11ac-2013 as the next high-throughput Wireless Local Area Network (WLAN) amendment. In this paper, we review the expected future WLAN scenarios and use-cases that justify the push for a new PHY/MAC IEEE 802.11 amendment. After that, we overview a set of new technical features that may be included in the IEEE 802.11ax-2019 amendment and describe both their advantages and drawbacks. Finally, we discuss some of the network-level functionalities that are required to fully improve the user experience in next-generation WLANs and note their relation with other on-going IEEE 802.11 amendments. 
\end{abstract}



\section{Introduction}

IEEE 802.11 Wireless Local Area Networks (WLANs) \cite{80211} are a cost-efficient solution for wireless Internet access that can satisfy most current communication requirements in domestic, public and business scenarios. 

Similar to other wireless technologies, WLANs have evolved by integrating the latest technological advances in the field as soon as they have become sufficiently mature, aiming to continuously improving the spectrum utilization and the raw WLAN performance. IEEE 802.11n-2009 adopted Single-user Multiple Input Multiple Output (SU-MIMO), channel bonding and packet aggregation. Those mechanisms were further extended in IEEE 802.11ac-2013, which also introduced Downlink Multi-user (MU) MIMO transmissions. In addition, new amendments such as the IEEE 802.11af-2013 and the IEEE 802.11ah-2016 are further expanding the application scenarios of WLANs, which include cognitive radio, long-range communication, advanced power saving mechanisms, and support for Machine to Machine (M2M) devices.

Partly because of their own success, next-generation WLANs face two main challenges. First, they must address dense scenarios, which is motivated by the continuous deployment of new Access Points (APs) to cover new areas and provide higher transmission rates. Second, the current evolution of Internet usage towards real-time high-definition audio and video content will also significantly increase users’ throughput needs in the upcoming years.

To address those challenges, the High-Efficiency WLAN (HEW) Task Group \cite{HEWTaskGroup} is currently working on a new high-throughput amendment named IEEE 802.11ax-2019. This new amendment will develop new physical (PHY) and medium access control (MAC) layer enhancements to further improve the WLAN performance, with a focus on the throughput and battery duration. This article overviews some of those new enhancements and describes the potential benefits and drawbacks of each one. We have grouped these enhancements into four main categories: spatial reuse, temporal efficiency, spectrum sharing and multiple-antenna technologies. Moreover, we also discuss several key system-level improvements for next-generation WLANs, as in addition to the IEEE 802.11ax-2019 amendment, they will likely implement other in-progress amendments such as IEEE 802.11aq-2016 (pre-association discovery of services), IEEE 802.11ak-2017 (bridged networks) and IEEE 802.11ai-2016 (fast initial link setup time) to satisfy the created expectations.


\section{Scenarios, Use-cases and Requirements} \label{Sec:scenarios}

The forecast number of devices and networks, the traffic characteristics and user demands for the 2020-2030 decade motivate the development of a new PHY/MAC IEEE 802.11 amendment to cope with the new challenges and usages WLANs will face \cite{HEWTaskGroup}. 

One of the most representative characteristics of WLANs is the use of Carrier Sense Multiple Access (CSMA/CA) as MAC protocol. It offers a reasonable trade-off between performance, robustness and implementation costs. Using CSMA/CA, when a node has a packet ready for transmission, it listens to the channel. Once the channel has been detected free (i.e., the energy level on the channel is lower than the CCA (Clear Channel Assessment) threshold, the node starts the backoff procedure by selecting a random initial value for the backoff counter. The node then starts decreasing the backoff counter while sensing the channel. Whenever a transmission, from either other nodes within the same WLAN or those belonging to other WLANs, is detected on the  channel, the backoff counter will be paused until the channel is detected free again, at which point the countdown is resumed. When the backoff counter reaches zero, the node starts transmitting. Figure \ref{Fig:CSMACA} shows an example of the CSMA/CA operation.    

\subsection{Dense WLAN scenarios}

\begin{figure}[t!!]
\centering
    \psfrag{AP}[][][0.7]{AP}
    \psfrag{STA}[][][0.7]{STA}
    \psfrag{DATA}[][][0.7]{DATA}
    \psfrag{ACK}[][][0.7]{ACK}

    \psfrag{successful}[][][0.8]{successful transmission}
    \psfrag{backoff}[][][0.8]{backoff countdown}
    \psfrag{busy}[][][0.8]{busy channel}
    \psfrag{collision}[][][0.8]{collision between two transmissions}

    \subfigure[Example of CSMA/CA temporal evolution with one AP and two STAs]{   \epsfig{file=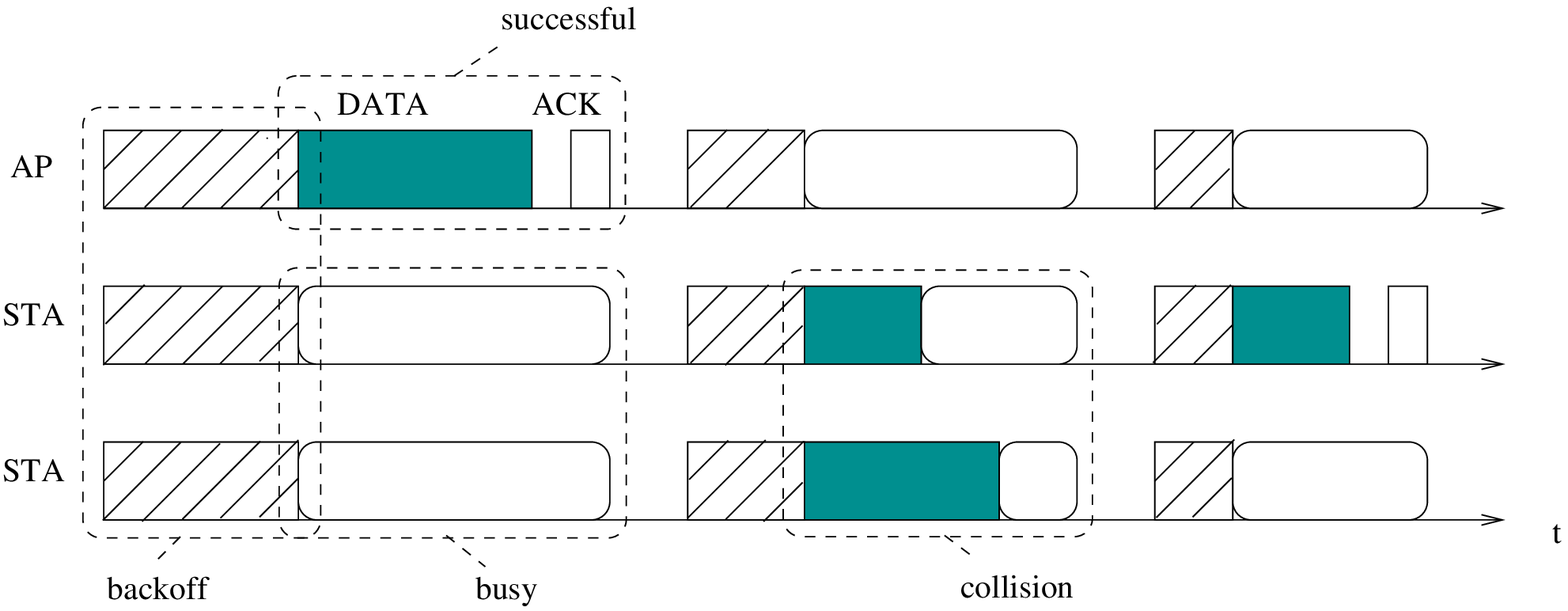,scale=0.65,angle=0}\label{Fig:CSMACA} }

    \vspace{1cm}
    \begin{tabular}{p{3.5cm}|c|c|p{8cm}}
    Scenario (Area, m$^2$)& APs & STAs & Description \\ 
    \hline \hline
    Stadium ($\sim$12500 m$^2$) & $>1000$  & $>50000$ &  Large events that require many APs to provide a satisfactory connectivity service able to support video uploading/downloading. \\ 
    \hline 
    Train  ($\sim$600 m$^2$) & $<10$  & $>1000$ & Full coverage inside a train to provide both work and entertainment services. \\
    \hline 
    Apartment building  - 4 floors, 6 apartments/floor - ($\sim$2400 m$^2$) & $>120 $ & $<480$ & Several short-range APs deployed in each apartment, offering full coverage and high data rates for bandwidth hungry entertainment applications, as well as connectivity for house appliances. Community APs may be also deployed in the corridors and shared spaces. \\ 
    \hline 
    \end{tabular}

    \psfrag{a}[][][1]{a) Stadium}
    \psfrag{b}[][][1]{b) Train}
    \psfrag{c}[][][1]{c) Floor of a building with several apartments}
    \psfrag{coverage}[][][0.8]{WLAN coverage area}
    \psfrag{SR-WLAN}[][][0.8]{Short-range WLAN}

    \vspace{1cm}
    \epsfig{file=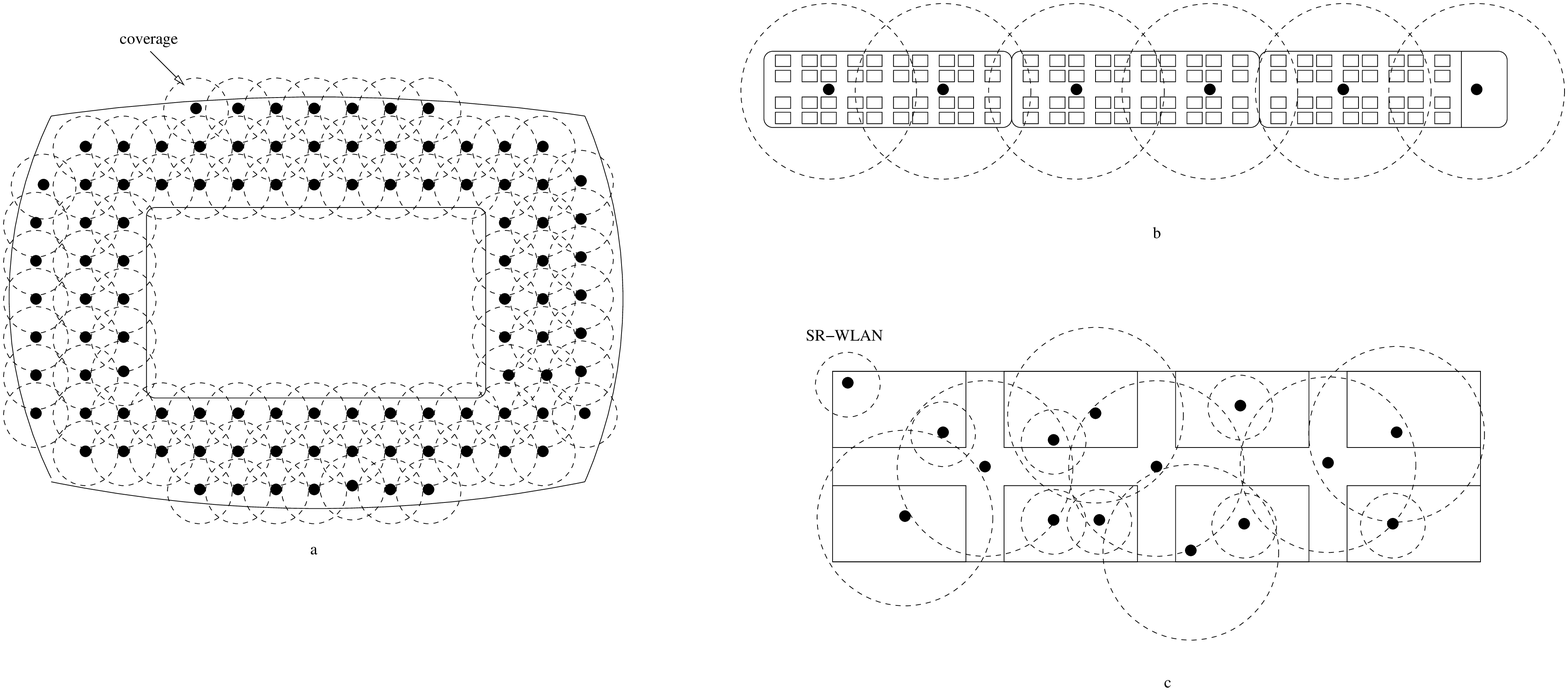,scale=0.4,angle=0}\\
    \caption{Key scenarios in next-generation WLANs.}
    \label{Fig:Scenarios}
\end{figure}

Providing high data rates in scenarios where the density of WLAN users is very high (e.g., 1 user/m$^2$) requires the deployment of many APs placed close to each other (e.g., within 5-10 m of one another). Figure \ref{Fig:Scenarios} depicts and describes three of those scenarios: a) a stadium b) a train, and c) an apartment building. In these dense scenarios, most relevant challenges are related to interference issues, which increase the packet error rate and reduce the number of concurrent transmissions in a given area by preventing neighboring WLANs from accessing the channel. Additionally, the presence of many stations (STAs) in the same area increases the chances that the backoff counters of two or more STAs reach zero simultaneously, which results in a collision.

In the stadium scenario, many people are concentrated in small areas because of a fair, a conference or a sporting event. The presence of many people results in a high density of STAs and the necessity for deploying many APs to offer satisfactory service. A fundamental challenge in these scenarios is to deploy, optimize and coordinate such a large number of APs and STAs.

Public transport is also a key scenario for next-generation WLANs because trains, buses and planes will offer broadband Internet access. In these scenarios, the user density may be notably high, with several people per square meter. Then, a smart AP coordination can help improve the spatial reuse, and the use of an efficient medium access protocol may help support many simultaneous contenders.

Finally, in the apartment building, we can find multiple autonomous and heterogeneous WLANs overlapping, including short-range WLANs that offer high transmission rates in small spaces \cite{bellaltachannel}. In this scenario, each WLAN is primarily configured independently of the others, where the channel selection, channel width and transmission power are randomly set or are simply the pre-set values. Therefore, autonomous WLANs must be able to implement smart decentralized self-configuration and self-adaption mechanisms to minimize the interference among them.

WLANs must also coexist with other wireless networks that operate in the ISM band, such as Wireless Sensor Networks and Personal Area Networks. In addition, Long-Term Evolution (LTE) operators currently consider deploying LTE networks in the ISM band \cite{abinader2014enabling}, which is known as LTE-Unlicensed, thus opening further coexistence challenges for WLANs. 


\subsection{Future WLAN usages}

Interactive and high-definition video applications are predicted to dominate future Internet usage. Two examples of applications that require throughputs of several Gbps are high-definition multi-party video conferences in business environments, which can help avoid unnecessary travel and meetings, and virtual reality entertainment applications at home, which include culture, films and games. Additionally, web surfing is moving further towards a multimedia experience, where rich text, images, audio and video content interact. Furthermore, file storage, management and synchronization in the cloud are becoming the standard in terms of content management and generation. Those applications are bandwidth-demanding and require both reliability and limited delay.


\subsection{Requirements}

Based on the aforementioned scenarios and expected use-cases, there are four key requirements for the IEEE 802.11ax-2019 amendment.

\begin{enumerate}
    \item \textbf{Coexistence}: WLANs operate as unlicensed devices in the ISM (Industrial, Scientific and Medical) bands. Therefore, the IEEE 802.11ax-2019 amendment has to include the required mechanisms to coexist both with the other wireless networks that also operate there and with the licensed devices.    
    \item \textbf{Higher throughput}: Improving both the system and user throughput requires the improved use of channel resources. IEEE 802.11ax-2019 aims for a 4-fold throughput increase compared with IEEE 802.11ac-2013. To achieve this goal, some new wireless technologies such as Dynamic CCA, OFDMA (Orthogonal Frequency Division Multiple Access), and advanced multiple-antenna techniques may be used.
    \item \textbf{Energy efficiency}: The target in IEEE 802.11ax-2019 is - at least - to not consume more than the previous amendments, considering the aforementioned 4-fold throughput increase, which requires both new low-power hardware architectures \cite{rajagopal2014} and new low-power PHY/MAC functionalities.
     \item \textbf{Backward Compatibility}: Because WLANs implementing IEEE 802.11ax-2019 must also support devices using any previous IEEE 802.11 PHY/MAC amendments, mechanisms must also be implemented to make it backward compatible (i.e., common frame headers and transmission rates), although it is a clear source of inefficiency. 
\end{enumerate}


\section{New Features and Concepts}  \label{Sec:features}

The IEEE 802.11ax-2019 amendment may include some new technical features compared with the IEEE 802.11ac-2013 amendment. We introduce them in this section, providing insight into their potential performance gains and limitations. All numerical results presented in this section are obtained using the analytical model and parameters from \cite{bellaltaChBondingInteractions}, unless otherwise is stated. 


\subsection{Spatial reuse} \label{Sec:SpatialCoexistence}

In dense scenarios, the combined use of CSMA/CA, a conservative CCA and a high transmit power level may result in scenarios with limited spatial reuse. A conservative configuration of both the CCA and transmit power levels minimizes the interference among the WLANs, which supports higher transmission rates. However, the number of concurrent transmissions is reduced, which may decrease the achievable area throughput. The alternatives that can be used to reach an optimal tradeoff between individual transmission rates and the number of concurrent transmissions that maximize the area throughput include adapting dynamically the transmit power level, the CCA level and the use of directional transmissions based on the observed network performance.

\begin{figure}[t!]
\centering
    
\psfrag{data range}[][][0.8]{Data range}
\psfrag{carrier sense range}[][][0.8]{Carrier sense range}
\psfrag{WLAN A}[][][0.7]{WLAN A}
\psfrag{WLAN B}[][][0.7]{WLAN B}
\psfrag{WLAN C}[][][0.7]{WLAN C}
\psfrag{channels}[][][0.6]{channels}

\subfigure[Three overlapping WLANs]{\epsfig{file=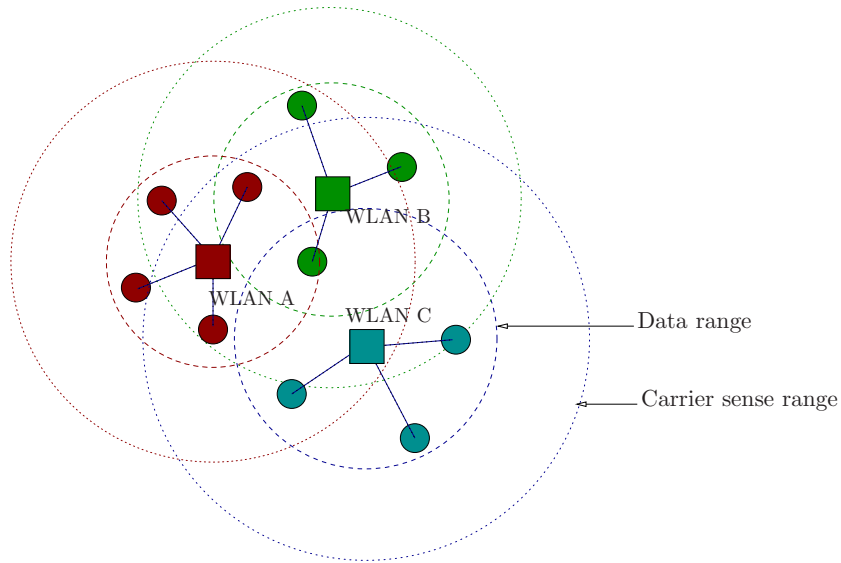,scale=0.45,angle=0}\label{Fig:TPC}}\\
\subfigure[Channel used by each WLAN]{\epsfig{file=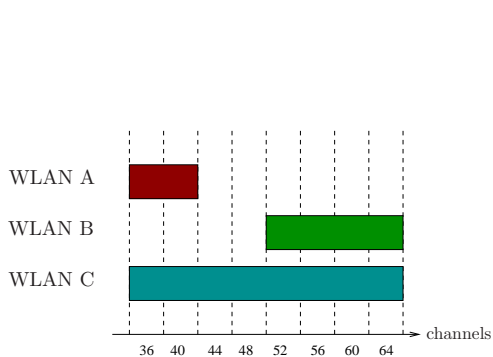,scale=0.45,angle=0}\label{Fig:ChAllocation}}\quad
\subfigure[Throughput achieved by each WLAN]{\epsfig{file=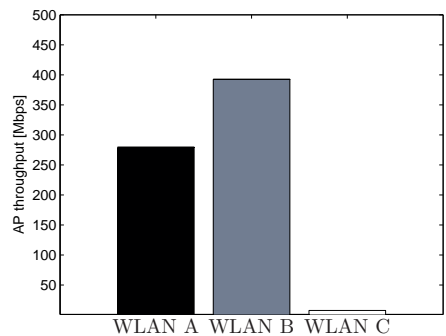,scale=0.45,angle=0}\label{Fig:Results-Overlapping}}
\caption{Throughput unfairness between overlapping WLANs.}
\end{figure}

Figure \ref{Fig:TPC} shows three neighboring WLANs. The channels that each WLAN uses are shown in Figure \ref{Fig:ChAllocation}. Because WLANs A and C, and B and C, partially share their channels, they overlap. The three APs are inside the carrier sense range of the others as shown in Figure 2(a), which pauses their backoff if either of the other two transmits. Although WLAN C uses the widest channel, it achieves the lowest throughput because it overlaps with WLANs A and B that are independent between them (Figure \ref{Fig:Results-Overlapping}).

\subsubsection{Dynamic adaptation of the transmit power and CCA levels} 

Reducing the used transmission power in a WLAN reduces its influence area, which benefits the spatial reuse. However, it may result in a larger number of packet errors and lower transmission rates, as well as to increase the number of hidden nodes. 

Alternatively, to reduce the area of influence of neighboring WLANs and increase each WLAN’s chances to transmit, the nodes in a WLAN may increase their CCA level, hence requiring a higher energy level in the channel to consider it as occupied and pause the backoff countdown. In \cite{jamil2014improving}, significant throughput gains are achieved by tuning the CCA level in a multi-cell WLAN scenario. The downside of increasing the CCA level is again the higher interference that a node may suffer, which could be detrimental in some cases.

\subsubsection{Beamforming}

Omnidirectional transmissions homogeneously spread the transmitted energy in all directions, which fills the channel with energy in areas where it is not required. Concentrating the energy towards the desired destination improves the spatial reuse because the devices that are placed in other directions will observe the channel as being empty and, therefore, start their own transmissions concurrently. However, similar to the previous case, those nodes outside the energy beam may also become hidden nodes.
    

\subsection{Temporal Efficiency} \label{Sec:TempEfficiency}

The backoff countdown, packet headers, interframe spaces, collisions and retransmissions are an intrinsic part of the CSMA/CA channel access scheme, but they significantly decrease the effective time that a node spends transmitting data every time it accesses the channel. IEEE 802.11ax-2019 may include several solutions to mitigate such overheads.

\subsubsection{Control Packets}

The time consumed by the exchange of control packets may result in large overheads, particularly because they are usually transmitted at a low rate. Common control packet exchanges between the AP and STAs include the RTS/CTS exchange to avoid hidden nodes and ACKs to acknowledge the reception of data packets. 

Additionally, some of the new technical features described in next sections that enable multi-user transmissions may require a frequent exchange of control packets to synchronize all involved STAs, hence also increasing the control packets overheads.

\subsubsection{Packet Headers, Aggregation \& Piggy-Backing}

Packet aggregation was introduced in IEEE 802.11n-2009 to reduce temporal overheads by combining short packets into a longer one. Using packet aggregation, multiple packets can be transmitted with a single backoff, DIFS, SIFS, PHY header and ACK. 

The packet header overheads can be reduced by supporting variable size headers and using only the minimum required fields for every packet. Additionally, the use of shorter identifiers instead of the full MAC address is considered. 

Moreover, the piggybacking of ACKs with DATA will improve the efficiency, although some changes in the current setting of the Network Allocation Vector (NAV) are required because the full transmission duration is unknown to the transmission initiator.

\subsubsection{Efficient retransmissions}

Packet errors are also a source of overhead because they currently require the full retransmission of the data packet. Further work about the use of incremental redundancy-based ARQs can reduce the time spent in retransmissions, although it implies some extra complexity in both transmitter and receiver firmware.

\subsubsection{Simultaneous Transmit and Receive}

By allowing the AP and a STA to simultaneously transmit and receive (STR), which is commonly known as full-duplex communication, the channel capacity can be theoretically doubled \cite{choi2010achieving}. Using CSMA/CA, the only way they can have full duplex communication is if they finish their backoff countdown simultaneously. Otherwise, one will start transmitting before the other, causing the latter to pause its backoff until the former finishes its transmission. Only when the number of active STAs is low, they have bidirectional and saturated traffic flows and use a small backoff contention window; here, the use of STR can result in significant gains. Therefore, specific channel access mechanisms should be considered in the IEEE 802.11ax-2019 amendment in case the STR capability is finally considered. Alternatively, the STR capability can allow WLANs to replace the Collision Avoidance (CA) feature of CSMA/CA with the Collision Detection (CD) one, because collisions can be promptly detected and resolved. 

\subsubsection{Collision-free MAC protocols}

Collisions represent an important waste of channel resources in WLANs. IEEE 802.11ax-2019 may consider enhancing or changing the underlying CSMA/CA protocol to minimize collisions. There are two possibilities: moving to a centralized solution or enhancing the current CSMA/CA protocol. Because centralized options such as the Hybrid coordination function Controlled Channel Access (HCCA) were never adopted in WLANs, a focus on enhancing the CSMA/CA appears more plausible. CSMA/ECA (CSMA with Enhanced Collision Avoidance) is a particularly good candidate to replace CSMA/CA because it is backward compatible, is easily implemented and outperforms CSMA/CA in all cases \cite{sanabria2013future}.

\begin{figure}[t!]
    \centering
    \psfrag{time}[][][0.8]{t}
    \psfrag{Deterministic backoff}[][][0.7]{Deterministic backoff}
    \psfrag{RTS}[][][0.7]{RTS}
    \psfrag{CTS}[][][0.7]{CTS}
    \psfrag{ACK}[][][0.7]{ACK}
    \psfrag{DATA}[][][0.7]{DATA}
    \psfrag{STR}[][][0.8]{Simultaneous transmissions using the STR capability}
    \psfrag{AP}[][][0.7]{AP}
    \psfrag{STA A}[][][0.7]{STA A}
    \psfrag{STA B}[][][0.7]{STA B}
    \psfrag{STA C}[][][0.7]{STA C}
    \epsfig{file=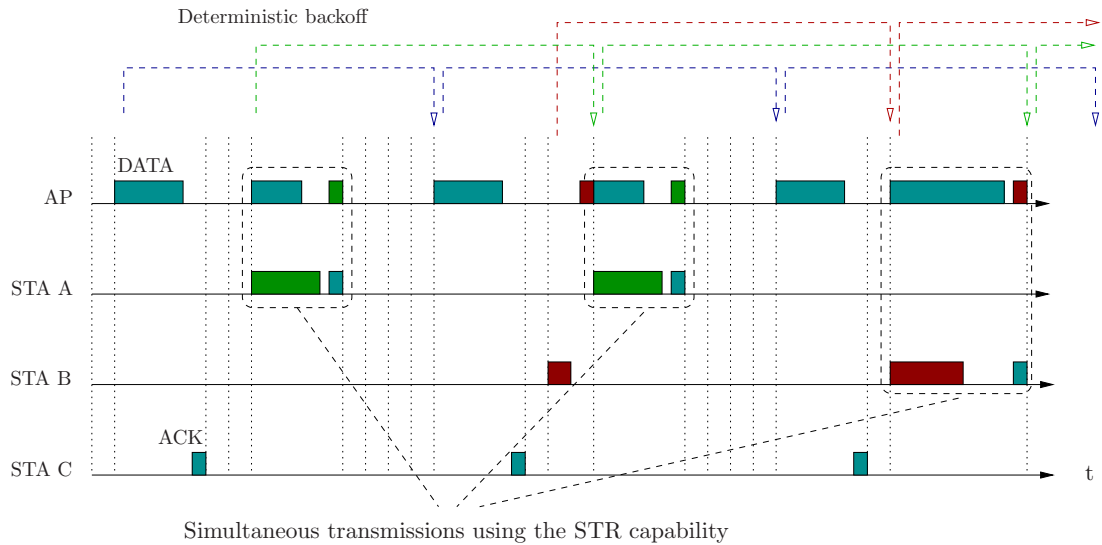,scale=0.60,angle=0}
    \caption{CSMA/ECA operation. It can be observed how the use of a deterministic backoff allows to predict when a node will transmit again after a successful transmission.}
\label{Fig:ECA}
\end{figure}

Figure \ref{Fig:ECA} shows the basic operation of CSMA/ECA with STR. Compared with CSMA/CA, the main difference observed for CSMA/ECA is its use of a deterministic backoff after successful transmissions. This deterministic backoff guarantees that after some time, a collision-free schedule can be achieved. In addition, because the AP can learn when the STAs will transmit, the use of STR can provide huge performance gains.

\subsection{Spectrum Sharing} \label{Sec:spectrum_sharing}

An unplanned deployment of WLANs results in a chaotic and fragmented spectrum occupancy, which causes many inefficiencies and undesirable interactions among neighboring WLANs \cite{bellaltaChBondingInteractions}. To improve the spectrum usage efficiency, two main approaches can be considered in IEEE 802.11ax-2019: dynamic channel bonding and OFDMA.

\subsubsection{Dynamic Channel Bonding}

To adapt to the instantaneous channel occupancy, IEEE 802.11ax-2019 may consider to extend the Dynamic Bandwidth Channel Access (DBCA) scheme introduced in the IEEE 802.11ac-2013 amendment \cite{park2011ieee}. Using DBCA, only the available channel width is used at each transmission, which allows the WLANs to adapt to the instantaneous spectrum occupancy. This mechanism helps fill most spectrum gaps and share them fairly among neighboring WLANs.

\subsubsection{OFDMA}

\begin{figure}[t!]
    \centering
    \psfrag{f}[][][0.8]{f}
    \psfrag{t}[][][0.8]{t}
    \psfrag{Band}[][][0.7]{Band}
    \psfrag{Occupied Spectrum}[][][0.7]{Occupied Spectrum}
    \psfrag{Free Spectrum}[][][0.7]{Free Spectrum}
    \psfrag{RTS}[][][0.7]{RTS'}
    \psfrag{CTS}[][][0.7]{CTS}
    \psfrag{ACK}[][][0.7]{ACK}
    \psfrag{DATA}[][][0.7]{DATA}
    \subfigure[Dynamic Channel Bonding and OFDMA]{\epsfig{file=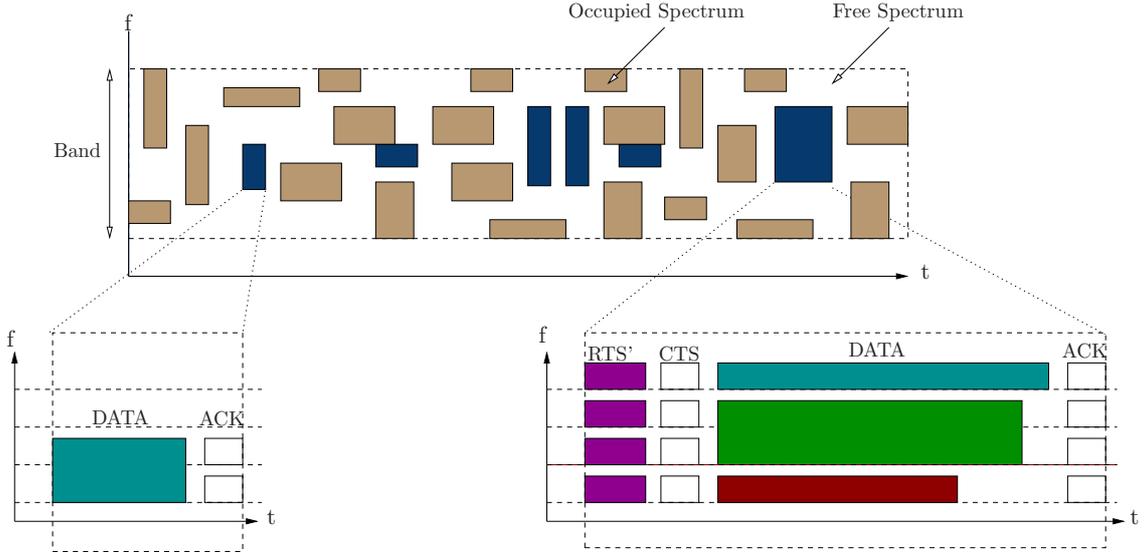,scale=0.5,angle=0}\label{Fig:DSA-Plot}}
    \subfigure[Achievable downlink throughput using a channel width of 160 MHz and OFDMA.  The RTS' size is $120+56\cdot N_{\text{tx}}$ bits, with $N_{\text{tx}}$ the number of OFDMA subchannels]{\epsfig{file=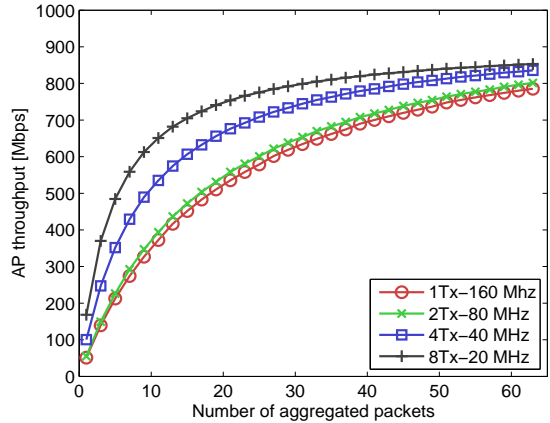,scale=0.55,angle=0}\label{Fig:DSA-Results}}
    \caption{Dynamic Spectrum Access with and without OFDMA.}
    \label{Fig:DSA}
\end{figure}

The use of OFDMA adds a new degree of flexibility to the use of spectrum resources by dividing the channel width into multiple narrow channels. Then, these narrow channels can be used to transmit to multiple users in parallel \cite{valentin2008integrating}. A basic implementation of OFDMA in WLANs may simply consider the use of multiple independent 20 MHz channels. This approach is shown in Figure \ref{Fig:DSA}: when channel bonding is used, each 20 MHz subchannel can be independently allocated to a different user. The RTS' packet has been extended to announce  the subchannels allocation to the STAs. Additionally, OFDMA may enable the use of non-contiguous channel bonding and remove the requirement to use only 20 MHz consecutive channels.

Figure \ref{Fig:DSA-Plot} shows an example where Dynamic Channel Bonding and OFDMA operate together. The upper part of Figure \ref{Fig:DSA-Plot} shows a snapshot of the spectrum occupancy for a group of neighboring WLANs. The lower part of Figure \ref{Fig:DSA-Plot} shows two transmissions: a node in the target WLAN transmits to a single user via a bonded channel of 40 MHz (left), and a node uses a bonded channel of 80 MHz and OFDMA to transmit to three different users (right). Figure \ref{Fig:DSA-Results} shows the AP throughput when OFDMA is used to split a 160 MHz channel into multiple subchannels. The parallelization of temporal overheads clearly improves the throughput.

\subsection{Multiple Antennas} \label{Sec:MultipleAntennas}

Spatial Multiplexing using multiple antennas at both AP and STAs remains one of the key technologies to achieve a high throughput in WLANs. IEEE 802.11ax-2019 will continue implementing both SU-MIMO and Downlink MU-MIMO, as in IEEE 802.11ac-2013. However, it may also include or provide support for Uplink MU-MIMO, Massive MIMO and Network MIMO, as well as to support distributed antenna solutions.

\subsubsection{Multi-user MIMO}

Multi-user MIMO enables multiple simultaneous transmissions to different STAs from the AP in the downlink, and from multiple STAs to the AP in the uplink. A survey of MU-MIMO MAC protocols for WLANs is presented in \cite{liao2014mu}, where the challenges and requirements to design a MU-MIMO MAC protocol are introduced, and several uplink and downlink MAC proposals are reviewed. 

In the downlink, a challenge for IEEE 802.11ax-2019 is to reduce the channel sounding overheads given the same explicit approach as in IEEE 802.11ac-2013 is considered. The overhead caused by the explicit channel sounding protocol implemented  depends on the channel sounding rate and number of sounded STAs, which can result in an unacceptable overhead in scenarios with many STAs. Solutions to reduce such a large overhead, apart from replacing the current channel sounding protocol with a more efficient solution, will require the use of smart schedulers that consider the current traffic patterns and the Quality of Service (QoS) demands from the users to decide when the CSI has to be requested, and from which STAs. 

To support MU-MIMO transmissions in the uplink, the IEEE 802.11ax-2019 amendment must also detail how the uplink CSI from each STA is obtained by the AP, introduce a mechanism to signal a group of STAs to simultaneously start a transmission, and include techniques to overcome channel calibration and timing issues in order to efficiently decode all the simultaneously received packets at the AP.   

In both cases, using the collected CSI, the AP has to select the specific STAs that will take part in next MU-MIMO transmission. Therefore, the design of an efficient mechanism to create groups of STAs with low spatial channel correlation and similar channel quality is still an open challenge. Failing on properly creating those groups may prevent next-generation WLANs to fully benefit from MU-MIMO technology.

\begin{figure}[t!]
    \centering
    \psfrag{f}[][][0.8]{f}
    \psfrag{t}[][][0.8]{t}
    \psfrag{AP}[][][0.9]{AP}
    \psfrag{STA A}[][][0.7]{STA A}
    \psfrag{STA B}[][][0.7]{STA B}
    \psfrag{STA C}[][][0.7]{STA C}
    \psfrag{PHY}[][][0.7]{PHY header}
    \psfrag{UPLINK}[][][0.7]{Uplink}
    \psfrag{DOWNLINK}[][][0.7]{Downlink}
    \psfrag{SUMIMO}[][][0.7]{SU-MIMO transmission}
    \psfrag{MUMIMO}[][][0.7]{MU-MIMO transmission}
    \psfrag{DATA transmission duration}[][][0.7]{DATA transmission duration}
    \psfrag{RTS''}[][][0.7]{RTS''}
    \psfrag{CTS}[][][0.7]{CTS}
    \psfrag{ACK}[][][0.7]{ACK}
    \psfrag{DATA}[][][0.7]{DATA}
    \subfigure[Downlink and Uplink MU-MIMO transmissions in WLANs]{\epsfig{file=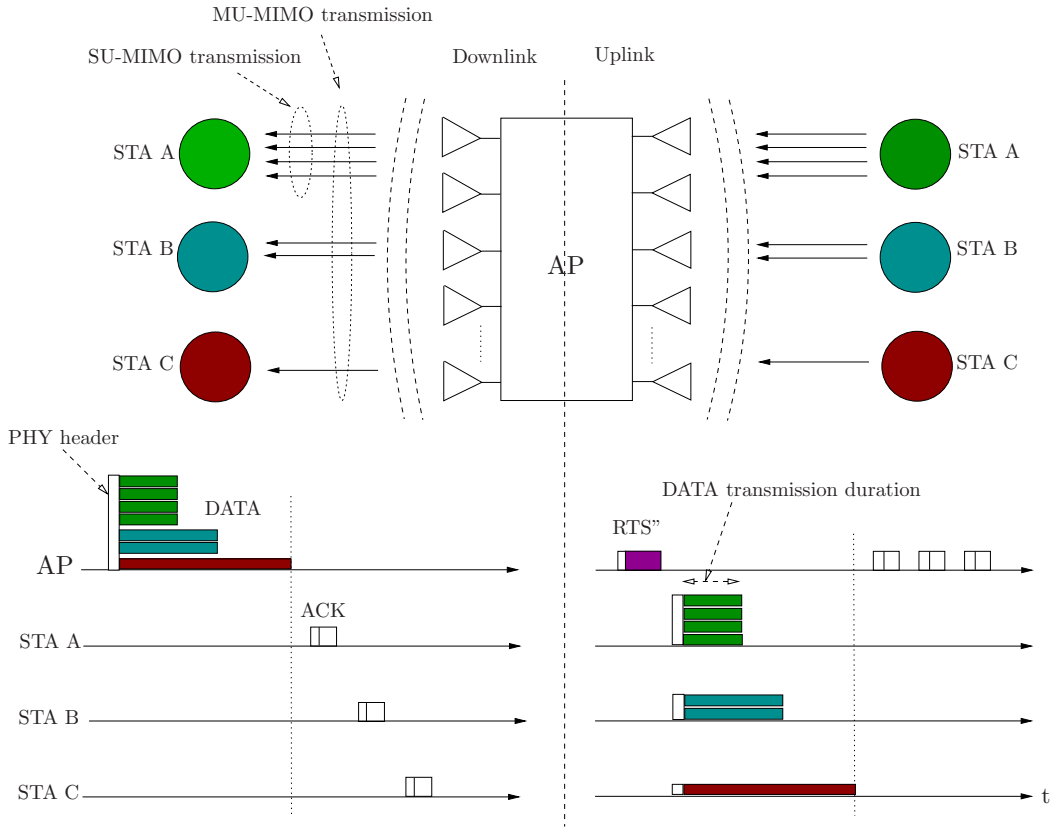,scale=0.50,angle=0}\label{Fig:MIMO}}
    \subfigure[Downlink Throughput. The AP is equipped with sixteen antennas. Each configuration is identified by the code x:y:z, where x refers to the total number of transmitted spatial streams, y to the number of destinations, and z to the number of streams transmitted to each destination]{\epsfig{file=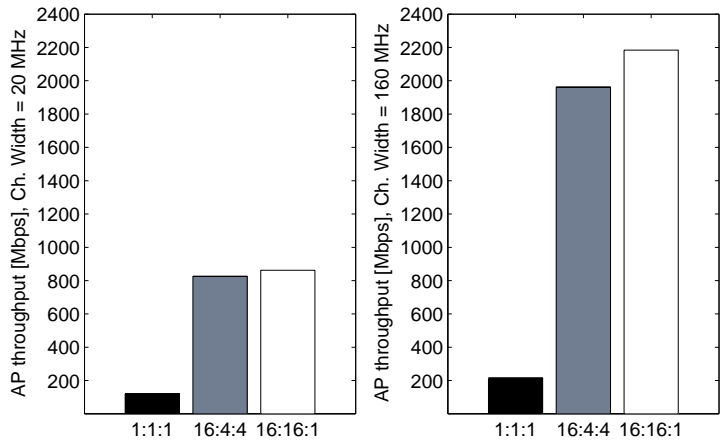,scale=0.6,angle=0}\label{Fig:MIMOChBonding}}
    \caption{Multi-user spatial multiplexing.}
\end{figure}

Figure \ref{Fig:MIMO} shows an AP with three STAs. In the left side, we have three downlink MU-MIMO transmissions. The ones directed to STAs A, B and C contain four, two and one SU-MIMO spatial streams, respectively. To start a downlink transmission, the AP omnidirectionally sends the PHY header with information about the group of selected STAs and the number of spatial streams that are transmitted to each STA in SU-MIMO mode. On the right side, we show an uplink MU-MIMO transmission. In IEEE 802.11ax-2019, because several STAs are unlikely to finish their backoff countdowns at the same time, uplink MU-MIMO transmissions may be started by the AP using a special RTS” packet containing information about the STAs that can transmit in parallel. Then, the selected STAs will simply start transmitting at the same time and wait to receive the corresponding ACKs. This approach allows to synchronize all selected STAs but requires the knowledge of the STAs's buffer occupancy by the AP, which can be provided by the same STAs in previous transmissions. Figure \ref{Fig:MIMOChBonding} shows the throughput achieved by the AP in two downlink MU-MIMO configurations (cases 16:4:4 and 16:16:1). The obtained throughput values are compared with the case where a single spatial stream is transmitted to only one destination (case 1:1:1).


\subsubsection{Massive MIMO}

Massive MIMO refers to the case where the AP has many more antennas than STAs and uses them to create a nearly identical number of point-to-point links as the number of active STAs \cite{shepard2012argos}. In addition to the cost of APs, the extra processing complexity and higher energy consumption, other open challenges for massive MIMO include obtaining the CSI information; WLANs may require switching to an implicit channel feedback approach.

\subsubsection{Network MIMO}

In coordinated WLAN deployments, network MIMO can be used to minimize the interference among simultaneous transmissions from different APs. The idea behind network MIMO is that different APs can coordinate the transmissions as if they were a large array of antennas, which reduces the inter-transmission interference and increases the spatial reuse \cite{zhang2013nemox}. However, effectively solving the tight synchronization requirements among the APs remains an open challenge.


\section{WLAN-Level Improvements} \label{Sec:network}

\begin{figure}[h!!!!!]
\centering
    \psfrag{IEEE 802.11ax}[][][0.7]{IEEE 802.11ax}
    \psfrag{IEEE 802.11ac}[][][0.7]{IEEE 802.11ac}
    \psfrag{IEEE 802.11ad}[][][0.7]{IEEE 802.11ad}
    \psfrag{IEEE 802.11ay}[][][0.7]{IEEE 802.11ay}
    \psfrag{IEEE 802.11ak}[][][0.7]{IEEE 802.11ak}
    \psfrag{IEEE 802.11aq}[][][0.7]{IEEE 802.11aq}
    \psfrag{IEEE 802.11ai}[][][0.7]{IEEE 802.11ai}
    \psfrag{IEEE 802.11n}[][][0.7]{IEEE 802.11n}
    \psfrag{IEEE 802.11ah}[][][0.7]{IEEE 802.11ah}    
    \psfrag{IEEE 802.11aa}[][][0.7]{IEEE 802.11aa}    

    \psfrag{PHY/MAC}[][][0.8]{PHY/MAC}    
    \psfrag{Extended }[][][0.8]{Extended}    
    \psfrag{Functionalities}[][][0.8]{Functionalities}

    \epsfig{file=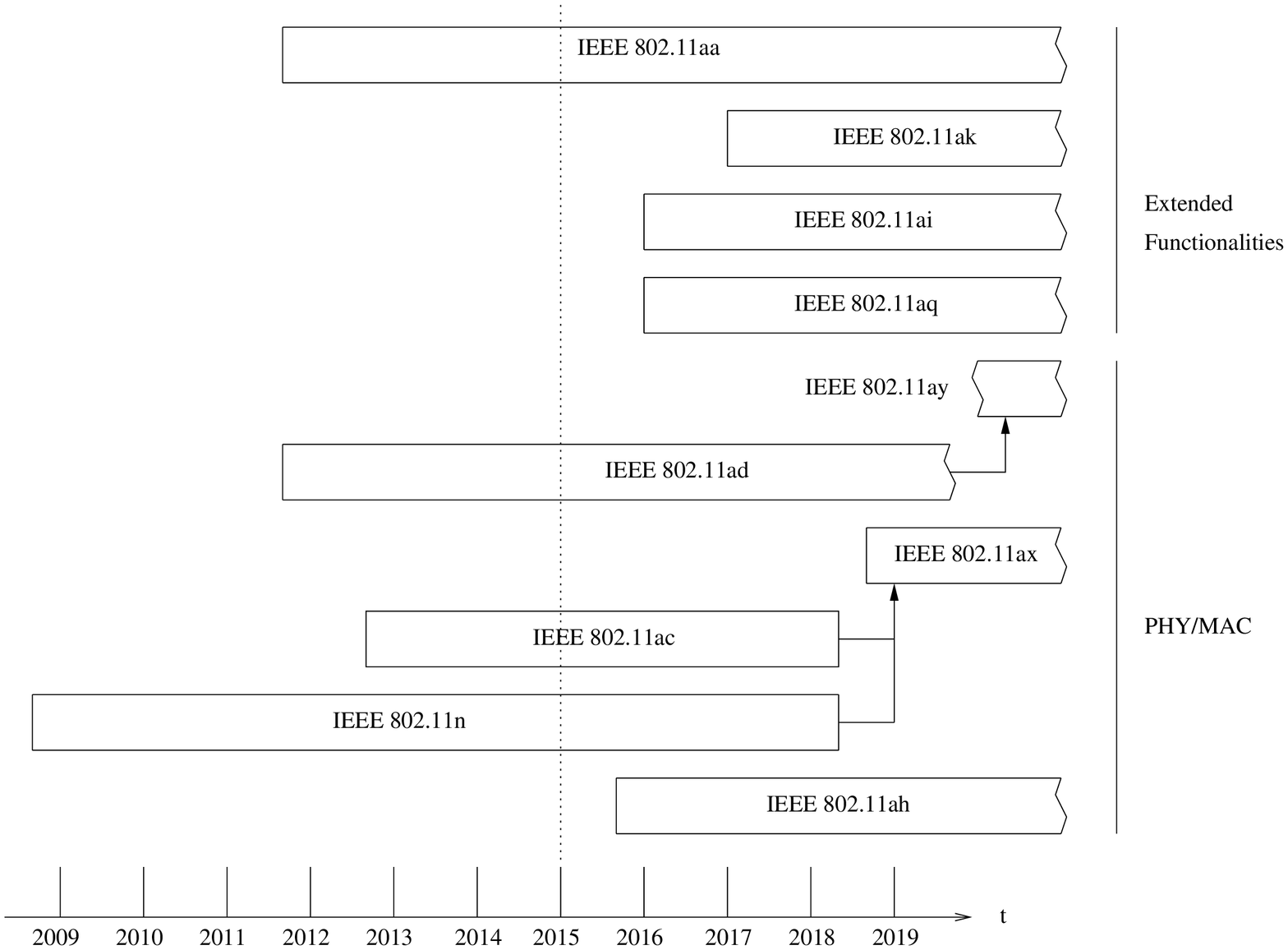,scale=0.65,angle=0}\\
    \vspace{0.5cm}

    \begin{tabular}{p{5cm}|p{3.5cm}|p{6cm}}
        Feature & Amendment & Objective\\
        \hline \hline
        Fast handoff between APs & IEEE 802.11ai-2016 & Handoff duration below 100 msecs by implementing preemptive channel sensing and user authentication. \\
        \hline        
        D2D/WI-FI Direct & WiFi Alliance feature & STAs with D2D capability can decide to create their own WLAN to communicate directly. \\
        \hline
        Multi-technology APs \& Bridging capabilities & IEEE 802.11ak-2017 & Interconnection of multiple WLANs.     \\
        \hline
        Video Traffic Differentiation & IEEE 802.11aa-2012 & Stream prioritization and groupcast mechanisms.\\
        \hline        
        AP/Service selection & IEEE 802.11aq-2018 & APs provide pre-association information about the services offered to help STAs to choose between multiple APs\\
        \hline
    \end{tabular}
    \caption{Upcoming IEEE 802.11 amendments.}\label{Fig:Amendments}
\end{figure}

The user experience in next-generation WLANs will not be simply enhanced by increasing the achievable network and user throughput as previous technical features do. To achieve that goal, apart from the IEEE 802.11ax-2019 amendment, there are other IEEE 802.11 amendments in progress. Figure \ref{Fig:Amendments} shows the most significant ones, including the new features they are targeting.

In scenarios where multiple APs are used to offer a large coverage area and higher data rates by deploying multiple WLANs nearby, a mechanism that enables a fast hand-off among the APs is required because the current large delays when a STA switches to a new AP are unacceptable. The IEEE 802.11ai-2016 amendment targets this challenge. It aims to provide a handoff duration below 100 msecs by implementing preemptive channel sensing and user authentication. Alternatives based on network virtualization and software defined networks are also an interesting option as they are able to centralize all decisions, including the handoff between overlapping APs. Probably, future IEEE amendments will consider such an approach.
  
Device-to-device (D2D) communications, which are also known as Wi-Fi Direct in the IEEE WLAN context, will also be a key element in next-generation WLANs \cite{camps2013device}. Examples of D2D communications are file synchronization with an external hard disk and instantaneous file exchange between a mobile phone and a projector/smart television. The use of D2D communications will reduce the amount of airtime required for each transmission by allowing higher data rates and avoiding the use of the AP as a relay. Beamforming is an interesting feature for D2D communication because it may allow concurrent transmissions inside the same WLAN by different groups of nodes. 

IEEE 802.11ax-2019 aims to operate in the 2.4 and 5 GHz bands, by superseding and integrating the IEEE 802.11n-2009 and IEEE 802.11ac-2013 amendments. In a few years, every single AP will most likely implement two IEEE 802.11ax-2019 instances that independently operate at 2.4 and 5 GHz, with a IEEE 802.11ah-2016 instance at 1 GHz for M2M and long-range communications and one IEEE 802.11ad-2012 (or the future IEEE 802.11ay-202x) at 60 GHz for very fast millimeter Wave communications. In this situation, the STAs of any of those networks should be able to communicate with the STAs in any of the other WLANs. This is in the aim of the IEEE 802.11ak-2017 amendment, which focuses on adding bridging capabilities to WLANs.

Finally, next-generation WLANs will use traffic differentiation, flow admission control and groupcast mechanisms from the IEEE 802.11e-2007, IEEE 802.11ae-2012 and IEEE 802.11aa-2012 amendments to support multimedia traffic with the required QoS. Also, further advances in Power Saving Mechanisms are expected to keep the WLAN energy consumption as low as possible.


\section{Conclusion}  \label{Sec:conclusions}

We have reviewed some technological options that could be included in the IEEE 802.11ax-2019 amendment for next-generation WLANs. Individually, all those solutions offer some performance gains by improving the spatial reuse and the spectrum utilization. However, the analysis of the actual performance gains when several of those solutions are combined and used simultaneously is still an open challenge requiring significant research efforts in the next years. For instance, for a given transmission power, multiplexing different users in a single transmission using channel bonding and spatial multiplexing reduces the received power per Hertz and user,  which may require the selection of lower order modulation schemes and coding rates, and the theoretical gain in throughput obtained by the combined use of spatial multiplexing and channel bonding may be lost by the larger transmission delay.

Finally, in addition to the higher achievable throughput provided by IEEE 802.11ax-2019 PHY/MAC features, we believe that the most disruptive performance and user experience improvements in next-generation WLANs will be also related to the development and implementation of the other recently approved or ongoing amendments, and by developing efficient and smart mechanisms to improve the coexistence and cooperation among WLANs.


\section*{Acknowledgements}

This work was partially supported by the Spanish and Catalan governments through the projects TEC2012-32354 and SGR-2014-1173 respectively.

\bibliographystyle{unsrt}
\bibliography{Bib_HEW2}

\begin{thebibliography}{10}

\bibitem{80211}
IEEE~802.11 WLANs.
\newblock {The Working Group for WLAN Standards}, 2015.

\bibitem{HEWTaskGroup}
{IEEE 802.11 Task Group AX}.
\newblock {Status of Project IEEE 802.11ax High Efficiency WLAN (HEW)}.
\newblock {\em Webpage, accessed July 2015. [Online]. Available:
  http://www.ieee802.org/11/Reports/tgax\_update.htm}, 2015.

\bibitem{bellaltachannel}
Boris Bellalta, Azadeh Faridi, Jaume Barcelo, Alessandro Checco, and Periklis
  Chatzimisios.
\newblock {Channel Bonding in Short-Range WLANs}.
\newblock In {\em 20th European Wireless Conference European Wireless. IEEE,
  2014, pp. 1–7.}, 2014.

\bibitem{abinader2014enabling}
FN~Abinader~Jr, Erika~PL Almeida, Fabiano~S Chaves, Andr{\'e}~M Cavalcante,
  Robson~D Vieira, Rafael~CD Paiva, Angilberto~M Sobrinho, Sayantan Choudhury,
  Esa Tuomaala, Klaus Doppler, et~al.
\newblock {Enabling the coexistence of LTE and Wi-Fi in unlicensed bands}.
\newblock {\em Communications Magazine, IEEE}, 52(11):54--61, 2014.

\bibitem{rajagopal2014}
Sridhar Rajagopal.
\newblock {Power efficiency: The next challenge for multi-gigabit-per-second
  Wi-Fi}.
\newblock {\em Communications Magazine, IEEE}, 52(11):40--45, 2014.

\bibitem{bellaltaChBondingInteractions}
Boris Bellalta, Alessandro Checco, Alessandro Zocca, and Jaume Barcelo.
\newblock {On the Interactions between Multiple Overlapping WLANs using Channel
  Bonding}.
\newblock {\em Vehicular Technology, IEEE Transactions on}, 2015.

\bibitem{jamil2014improving}
Imad Jamil, Laurent Cariou, and Jean-Francois Helard.
\newblock {Improving the capacity of future IEEE 802.11 high efficiency WLANs}.
\newblock In {\em Telecommunications (ICT), 2014 21st International Conference
  on}, pages 303--307. IEEE, 2014.

\bibitem{choi2010achieving}
Jung~Il Choi, Mayank Jain, Kannan Srinivasan, Phil Levis, and Sachin Katti.
\newblock {Achieving single channel, full duplex wireless communication}.
\newblock In {\em Proceedings of the Sixteenth Annual International Conference
  on Mobile Computing and Networking}, pages 1--12. ACM, 2010.

\bibitem{sanabria2013future}
Luis Sanabria-Russo, Azadeh Faridi, Boris Bellalta, Jaume Barcelo, and Miquel
  Oliver.
\newblock Future evolution of csma protocols for the ieee 802.11 standard.
\newblock In {\em Communications Workshops (ICC), 2013 IEEE International
  Conference on}, pages 1274--1279. IEEE, 2013.

\bibitem{park2011ieee}
Minyoung Park.
\newblock {IEEE 802.11ac: Dynamic bandwidth channel access}.
\newblock In {\em Communications (ICC), 2011 IEEE International Conference on},
  pages 1--5. IEEE, 2011.

\bibitem{valentin2008integrating}
Stefan Valentin, Thomas Freitag, and Holger Karl.
\newblock {Integrating multiuser dynamic ofdma into ieee 802.11 wlans-llc/mac
  extensions and system performance}.
\newblock In {\em Communications, 2008. ICC'08. IEEE International Conference
  on}, pages 3328--3334. IEEE, 2008.

\bibitem{liao2014mu}
Ruizhi Liao, Boris Bellalta, Miquel Oliver, and Zhisheng Niu.
\newblock {MU-MIMO MAC Protocols for Wireless Local Area Networks: A Survey}.
\newblock {\em IEEE Communications Surveys and Tutorials.}, 2014.

\bibitem{shepard2012argos}
Clayton Shepard, Hang Yu, Narendra Anand, Erran Li, Thomas Marzetta, Richard
  Yang, and Lin Zhong.
\newblock {Argos: Practical many-antenna base stations}.
\newblock In {\em Proceedings of the 18th Annual International Conference on
  Mobile Computing and Networking}, pages 53--64. ACM, 2012.

\bibitem{zhang2013nemox}
Xinyu Zhang, Karthikeyan Sundaresan, Mohammad A~Amir Khojastepour, Sampath
  Rangarajan, and Kang~G Shin.
\newblock {NEMOx: scalable network MIMO for wireless networks}.
\newblock In {\em Proceedings of the 19th annual International Conference on
  Mobile Computing \& Networking}, pages 453--464. ACM, 2013.

\bibitem{camps2013device}
Daniel Camps-Mur, Andres Garcia-Saavedra, and Pablo Serrano.
\newblock {Device-to-device Communications with Wi-Fi Direct: overview and
  experimentation}.
\newblock {\em Wireless Communications, IEEE}, 20(3), 2013.

\end{thebibliography}


\end{document}